\documentclass[12pt]{article}
\usepackage{amssymb}

\input{tcilatex}

\begin{document}

\title{Birth of a Closed Universe of Negative Spatial Curvature}
\author{H. V. Fagundes and S. S. e Costa\thanks{%
Instituto de F\'{i}sica Te\'{o}rica, Universidade Estadual Paulista, S\~{a}o
Paulo, SP, Brazil. E-mail: helio@ift.unesp.br}}
\maketitle

\begin{abstract}
We propose a modified form of the spontanteous birth of the universe by
quantum tunneling. It proceeds through topology change and inflation, to
eventually become a universe with closed spatial sections of negative
spatial curvature and nontrivial global topology.
\end{abstract}

\section{INTRODUCTION}

The idea of spontaneous birth of the universe by quantum tunneling from a de
Sitter instanton has been coupled by Vilenkin \cite{vilenkin} to the
beginning of the inflationary expansion in a de Sitter spacetime with space
sections of positive curvature. We propose an extension of this process,
through topology change and inflation, to arrive at a Friedmann model with
closed hyperbolic spaces (CHS) as spatial sections. We are motivated by
previous work by Gott\cite{gott1} and Bucher et al. \cite{bucher} in the
matter of inflation leading to an $\Omega $$_{0}<1$ universe, but assume a
mechanism different from theirs, relying on De Lorenci et al.'s \cite
{lorenci} formalism to quantify the probability of changes in cosmic global
topology. There have been suspicions of incompatibility of CHS universes
with maps of the cosmic microwave background (CMB)\cite{pogosyan}; we refer
to a recent paper by Cornish et al. \cite{cornish} to answer to this
criticism.

\section{CREATION AND EVOLUTION}

We use Planck units $c=G=\hbar =1.$

De Sitter instanton has the metric (cf. \cite{vilenkin})

\begin{equation}
ds^{2}=dt^{\prime 2}+r_{0}^{2}\cos {}^{2}(t^{\prime }/r_{0})(d\chi ^{2}+\sin
{}^{2}\chi \ d\Omega ^{2}),  \label{metricInstanton}
\end{equation}
where $t^{\prime }=-it$, with $t$ imaginary time, $d\Omega ^{2}=d\theta
^{2}+\sin {}^{2}\theta \ d\varphi ^{2}$, $(\chi ,\theta ,\varphi )$ are
spherical coordinates, and $r_{0}$ is a constant of the order of Planck's
length or time. But instead of being defined on a four-sphere $S^{4}$, our
gravitational instanton has a more general topology: note that (i) the $%
S^{4} $ instanton is obtained by analytic continuation into imaginary time
of de Sitter spacetime with topology $R\times S^{3}$, where the real line $R$
is the time axis and the three-sphere $S^{3}$ is the spatial section \cite
{GandH}; (ii) de Sitter's metric of positive spatial curvature can be
assigned to any topology of the form $R\times M$, where $M=S^{3}/\Gamma $ is
the quotient space of $S^{3}$ by a discrete group of isometries $\Gamma ,$
which acts on $S^{3}$ without fixed points - see, for example, Ellis \cite
{ellis}, or Lachi\`{e}ze-Rey and Luminet \cite{LaLu}; and (iii) the same
process of analytic continuation into imaginary time can be done on these
new spacetime topologies. So we assume as a generalization of de Sitter's
instanton the metric of Eq. \ref{metricInstanton} defined on the quotient
spaces $S^{4}/\Gamma $ (which may not be manifolds but rather orbifolds -
see \cite{scott}, \S 2). The volume of $M,$ with its curvature normalized to
unity, is $2\pi ^{2}/($order of $\Gamma );$ its fundamental group $\pi
_{1}(M)$ is isomorphic to $\Gamma ,$ hence, except for the case $M=S^{3}$
(trivial $\Gamma ),M$ is multiply connected.

When this instanton becomes a real de Sitter spacetime, we get the
inflationary metric of positive spatial curvature (cf. \cite{vilenkin}, \cite
{gott1}), starting at $t=0,$ 
\begin{equation}  \label{deSitterSph}
ds^2=-dt^2+r_0\cosh {}^2(t/r_0)(d\chi ^2+\sin {}^2\chi \ d\Omega ^2),
\end{equation}
and the topology of the direct product $R\times M$, as implied above. The
parameter $r_0\,$ is related in \cite{gott1} to a constant energy density,
which is now interpreted as the initial value of the inflaton potential $%
V(\phi )$. As in \cite{bucher}, during this epoch the inflaton field $\phi $
is stuck in a false vacuum (region $A$ in Fig. 1), causing a growth that
erases inhomogeneities in $M$. But in our case this stage is very short. In
the numerical example of next Section, $t_f=r_0$ is the endtime of this
epoch, which therefore may hardly be called inflationary; see below.

At $t=t_{f}$ we assume that a topology and metric transition takes place;
the latter becomes

\begin{equation}
ds^{2}=-d\tau ^{2}+r_{0}^{2}\sinh {}^{2}(\tau /r_{0})(d\chi ^{\prime
2}+\sinh {}^{2}\chi ^{\prime }d\Omega ^{2}),  \label{deSitterHyp}
\end{equation}
with initial $\tau =\tau _{i}$ to be determined below. This is similar to
the metric change in \cite{gott1}, but here we shall assume it to happen as
a quantum transition in minisuperspace. The change in metric in a closed
space implies a change in topology, since the spherical space $S^{3}/\Gamma $
cannot support the hyperbolic metric in the spatial part of Eq. \ref
{deSitterHyp} -- cf. \cite{scott}, Theorem 5.2. This metric and topology
change is signaled by the small bump in the potential $V(\phi )$ -- region $B
$ in Fig. 1; we hope a mechanism for this process can be adapted from the
one recently developed by De Lorenci et al. \cite{lorenci}, who estimated
the probabilities of a few similar transitions. Their results, interpreted
with some liberty, indicate that a spherical cosmology is unstable against
quantum transitions, and has a good chance of becoming hyperbolic, as in our
case.

Here the spatial section becomes the CHS $M^{\prime }=H^3/\Gamma ^{\prime },$
where $H^3$ is hyperbolic three-space, and $\Gamma ^{\prime }\cong \pi
_1(M^{\prime })$ is a discrete group of isometries acting on $H^3$ without
fixed points -- cf. \cite{LaLu}. Spacetime topology becomes $R\times
M^{\prime }.$ Since this is a quantum process, we need not demand continuity
in the five-dimensional pseudoeuclidean space where de Sitter spacetime is
imbedded, as done in \cite{gott1}. But, as a working hypothesis, we
postulate that energy, hence physical volume, is conserved in the
transition, and we expect the change in the expansion factor to be small;
therefore the normalized volumes of $M$ and $M^{\prime }$ should be of the
same order of magnitude. This can be arranged, and $\tau _i$ calculated, as
will be seen in the next Section.

When space becomes $M^{\prime }$ it may again have density irregularites.
(To see this, imagine a uniform distribution of a thin film of matter over a
two-sphere, which suddenly gets a handle and becomes a torus; the film takes
a time to spread itself evenly over the new surface.) In the light of this,
it is not so essential that the previous stage gets homogenized, as it is
that its duration be short so that $M$ will not grow too much and $M^{\prime
}$ may become smooth; see next Section. At this point one might wonder why
is stage $M$ necessary at all. Perhaps because there is no gravitational
instanton \cite{hawking} that would actualize directly as a closed
hyperbolic universe.

Thus $M^{\prime }$ is neither equivalent to the nucleated bubble in \cite
{gott1} or \cite{bucher}, nor to the smooth patch in Kolb and Turner's \cite
{KandT} basic picture of inflation, for either of these is already
homogeneous as it appears on the scene. Besides, $M^{\prime }$ need not grow
to encompass the observable universe; when it does not, as in the example
below, the model predicts observable effects of the nontrivial topology. The
most obvious of these, but still difficult to verify, is the production of
multiple images of each source, so as to mimic the uniform distribution of
an open Friedmann model. See \cite{hvf1983};\cite{LaLu} and references
therein.

In \cite{bucher} this is the time of slow roll inflation, with $V(\phi )$
sloping as $-\mu ^3\phi $ towards the true vacuum. For simplicity here we
prefer the potential in \cite{gott1}, which has a plateau (region $C$ in
Fig. 1) of about the same height as in the false vacuum, so that inflation
proceeds at the same rate as before. This epoch is $\tau _i\leq \tau \leq
\tau _1$, with $\tau _1$ being determined by a continuity condition with the
next phase.

Finally \cite{gott1} a phase transition is arranged, the equation of state
going from $p=-\rho $ to $p=\rho /3,$ which ends the inflationary period.
There is reheating around the true vacuum ($D$ in Fig. 1), and spacetime
gets the usual Friedmann metric of negative spatial curvature

\begin{equation}  \label{FriedmannHyp}
ds^2=-d\tau ^2+a^2(\tau )(d\chi ^{\prime 2}+\sinh {}^2\chi ^{\prime }d\Omega
^2).
\end{equation}
Thus we reach the epoch of standard cosmology, except for the effects of the
compactness and multiple connectedness of the spatial sections -- see \cite
{LaLu}.

\section{NUMBERS}

A convenient family of candidates for $M=S^3/\Gamma $ are the lens spaces $%
L(p,q),$ where $p,q$ are coprime integers with $1\leq q\leq p/2$ \cite{SandT}%
. Their fundamental group, hence also $\Gamma ,$ is of order $p$, so their
volume is $2\pi ^2/p.$ Hyperbolic manifolds $M^{\prime }$ are known to exist
with normalized volumes from $0.94$ up \cite{weeks}.

As an example, let us take $M=L(50,1),$ with normalized volume $%
v_{sph}=0.394784$, and as $M^{\prime }$ the smallest known CHS,
Weeks-Matveev-Fomenko manifold (cf. \cite{hvf1996}), with $v_{hyp}=0.942707.$
Then the universe's largest half-diameter, $R_{\max }(t)=(\pi /4)r_0\cosh
(t/r_0),$ is smaller than the radius of the event horizon, $R_H(t)=2r_0\cosh
(t/r_0)\{\tan {}^{-1}[\exp (t/r_0)]-\pi /4\}$ for $t>0.8814r_0.$ Let us take 
$t_f=r_0$ as the endtime of this epoch, which is enough to homogenize $M$.
By the conservation of physical volume, $v_{sph}\cosh {}^3(1)=v_{hyp}\sinh
{}^3(\tau _i/r_0),$ whence $\tau _i=0.9865r_0.$

To find $\tau _{1}$, first we have to make sure that there is enough time
for inflation to smoothen an initial inhomogeneity in $M^{\prime }$. The
circumscribing radius of the Dirichlet domain given in \cite{weeks} is $%
r_{\max }=0.752470$ in comoving (or normalized) units; the maximal distance
between points in $M%
{\acute{}}%
$ is of this order of magnitude. The horizon's radius is $r_{H}(\tau )=\ln
[\tanh (\tau /2r_{0})/\tanh (\tau _{i}/2r_{0})].$ We can have $r_{H}(\tau
)\gtrsim r_{\max }$ only if $\tanh (\tau _{i}/2r_{0})<\exp (-r_{\max }),$ or 
$\tau _{i}<1.0232r_{0}.$ With $\tau _{i}$ as obtained above, we must have $%
\tau _{1}>4.17r_{0}.$ This value is compatible with the one we now obtain
from the continuity of the expansion factor, $r_{0}\sinh (\tau
_{1}/r_{0})=a(\tau _{1}).$ In the beginning of the Friedmann radiation era
that follows reheating, $a(\tau )=(2b_{\ast }\tau )^{1/2}$ (cf. \cite
{ohanian}), where $b_{\ast }=(8\pi G\rho _{rad,0}/3c^{2})^{1/2}a_{0}^{2},$
with $\rho $$_{rad,0}=$ present density of radiation energy $=4.6477\times
10^{-34}$ g\ cm$^{-3}$ \cite{KandT}, and $a_{0}=$ present value of $a(\tau )$%
. Taking $r_{0}=1.6160\times 10^{-33}$ cm (Planck's length), $\Omega
_{0}=0.3,$ and Hubble's constant $H_{0}=65$ km s$^{-1}$Mpc$^{-1}$, the
continuity condition gives $\tau _{1}=71.1r_{0},$ close to the value $%
69r_{0} $ in \cite{gott1}.

Today's physical volume of $M^{\prime }$ would be $4.64\times 10^{84}$ cm$^3$%
, while the observable universe -- interpreted as the region of repeated
cosmic images -- is about 200 times larger.

\section{REMARKS}

In the above reasoning the dynamics of spontaneous birth and inflation
processes in the modified scenario was presumed to be adaptable from
previous results. We plan to elaborate on this point, as also on the matter
of topology changes, and to present a more detailed picture in the future.
Thermodynamics considerations, prevalent in \cite{gott1} and not touched
upon here, may also be addressed in the wider study.

The question of the influence of space closure on the generation and growth
of primordial fluctuations has been left out, to be examined elsewhere. Here
we only comment on doubts that have appeared on whether a closed hyperbolic
universe would be compatible with the long wavelength modes of the CMB --
see \cite{pogosyan}. As pointed out in \cite{cornish}, in hyperbolic
universes there is no low cutoff for these modes. The confusion seems to
arise from mistaking the separations between equivalent points in the space
of images $H^3$ with maximal wavelengths of perturbations; but the latter
can be spread over closed geodesics, which may form knotted patterns of
increasing length inside the CHS. See also \cite{roukema}, \cite{FandG}.
Therefore these models are probably compatible with the spotted maps of the
CMB obtained by NASA's COBE satellite, but of course more research is needed
in this direction.

One of us (S. S. C.) thanks Funda\c c\~ao de Amparo \`a Pesquisa do Estado
de S\~ao Paulo (FAPESP) for a doctoral scholarship. H. V. F. is grateful to
Ruben Aldrovandi for conversations on geometry, and to Conselho Nacional de
Desenvolvimento Cient\'{\i}fico e Tecnol\'ogico (CNPq - Brazil) for partial
financial support.

\newpage\ \ 

Figure caption:\vspace{1.0in}

Fig. 1. This is a qualitative plot of the potential $V(\phi )$. $A$ is the
region of transient spherical topology, $B$ is where the tunneling to
hyperbolic space happens, $C$ is the region of most inflation, and $D$ is
where reheating takes place around the true vacuum.

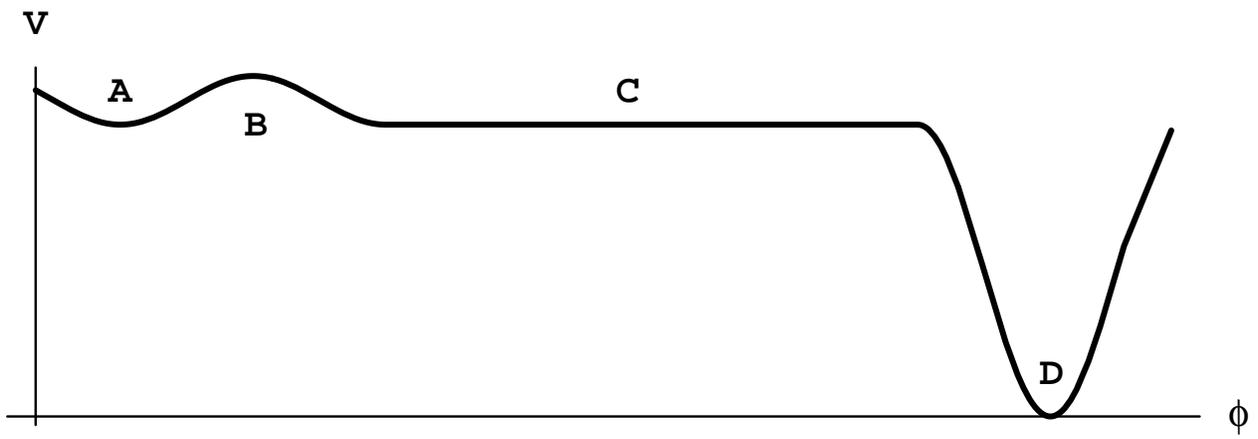
\end{document}